\documentclass[conference]{IEEEtran}
\IEEEoverridecommandlockouts
\usepackage{cite}
\usepackage{amsmath,amssymb,amsfonts}
\usepackage{algorithm,algorithmic}
\makeatletter
\newcommand\fs@norules{\def\@fs@cfont{\bfseries}\let\@fs@capt\floatc@ruled
  \def\@fs@pre{}%
  \def\@fs@post{}%
  \def\@fs@mid{\kern3pt}%
  \let\@fs@iftopcapt\iftrue}
\makeatother
\floatstyle{norules}
\restylefloat{algorithm}
\usepackage{graphicx}
\usepackage{multirow}
\usepackage{multicol}
\usepackage{booktabs}
\usepackage{adjustbox}
\usepackage{textcomp}
\usepackage{xcolor}
\usepackage{xspace}
\usepackage{tikz}
\usepackage[hidelinks]{hyperref}

\def\eg{\textit{e.g.}\@\xspace} 
\def\ie{\textit{i.e.}\@\xspace}

\def\wrt{w.r.t.\@\xspace}
\def\etal{\textit{et al.}\@\xspace} 

\title{AMB-FHE: Adaptive Multi-biometric Fusion with Fully Homomorphic Encryption}
\author{\IEEEauthorblockN{Florian Bayer and Christian Rathgeb}
\IEEEauthorblockA{\textit{da/sec–Biometrics and Internet-Security Research Group} \\
\textit{Hochschule Darmstadt}\\
Darmstadt, Germany \\
\{firstname\}.\{lastname\}@h-da.de}
}

\newcommand\copyrighttext{%
  \footnotesize \textcopyright 2025 IEEE. Personal use of this material is permitted.
  Permission from IEEE must be obtained for all other uses, in any current or future 
  media, including reprinting/republishing this material for advertising or promotional 
  purposes, creating new collective works, for resale or redistribution to servers or 
  lists, or reuse of any copyrighted component of this work in other works.
  }
\newcommand\copyrightnotice{%
\begin{tikzpicture}[remember picture,overlay]
\node[anchor=south,yshift=10pt] at (current page.south) {\fbox{\parbox{\dimexpr\textwidth-\fboxsep-\fboxrule\relax}{\copyrighttext}}};
\end{tikzpicture}%
}

\begin{document}
\maketitle
\copyrightnotice

\begin{abstract}
Biometric systems strive to balance security and usability. The use of multi-biometric systems combining multiple biometric modalities is usually recommended for high-security applications. However, the presentation of multiple biometric modalities can impair the user-friendliness of the overall system and might not be necessary in all cases. 

In this work, we present a simple but flexible approach to increase the privacy protection of homomorphically encrypted multi-biometric reference templates while enabling adaptation to security requirements at run-time: An adaptive multi-biometric fusion with fully homomorphic encryption (AMB-FHE). AMB-FHE is benchmarked against a bimodal biometric database consisting of the CASIA iris and MCYT fingerprint datasets using deep neural networks for feature extraction. Our contribution is easy to implement and increases the flexibility of biometric authentication while offering increased privacy protection through joint encryption of templates from multiple modalities.
\end{abstract}

\section{Introduction}
Biometric technologies are ubiquitous and increasingly used with mobile devices that are equipped with biometric capture devices to acquire samples from their users. Biometric authentication is often the preferred method of authentication for most users because it is convenient and provides adequate security for most use cases. Since a single biometric modality may lack sufficient effective entropy for higher security requirements,  multi-biometric authentication schemes are recommended for high-security applications. However, biometric data are considered sensitive and subject to governmental regulations\cite{GDPR} and international standards such as\cite{ISO-IEC-24745-TemplateProtection-2022} and therefore must be protected. 

During the past decade, homomorphic encryption (HE) has emerged as a highly attractive option for the protection of biometric data. This is due to several favorable attributes, such as (i) being able to perform function evaluation without decryption, (ii) being asymmetric, thereby solving the key distribution problem present with symmetric cryptography, and (iii) the underlying lattice problem is deemed post-quantum secure. However, HE schemes come with limitations that pose difficulties in practical applications: (i) computational overhead, and (ii) restrictions regarding data types they operate on. Computational efficiency gains are therefore well appreciated. For the envisioned application, we are concerned with a biometric \emph{verification} scenario, that is, an authentication request associated with a claimed identity, therefore constituting a 1:1 comparison operation. Nevertheless, reducing comparison time, computational overhead, and, even more importantly, saving the user inconvenient modality presentations is desirable to improve usability.

\begin{figure}
    \centering
    \includegraphics[width=1\linewidth]{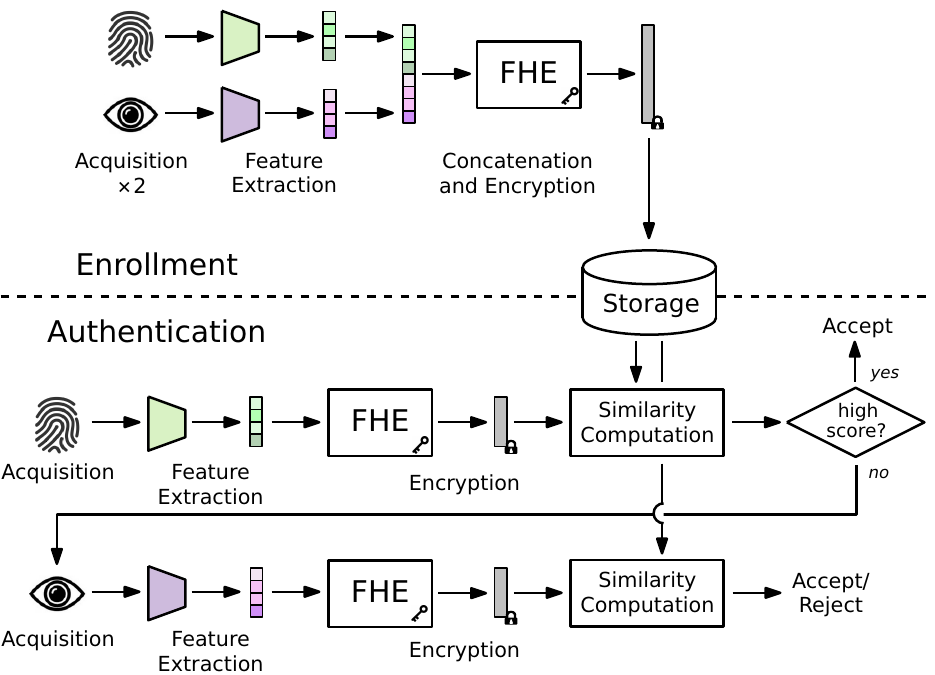}
    \caption{High-level overview of the AMB-FHE protocol for enrollment and verification with two biometric modalities. Upon enrollment, all supported modalities undergo a series of processes on the client device, including capturing, feature extraction, concatenation, and encryption. During authentication, a previously defined first modality is presented, and further modalities are captured sequentially if and only if the score falls below a certain security threshold.}
    \label{fig:overview}
\end{figure}
Due to increased overhead of longer cryptographic keys and performance gains through larger ciphertext sizes, application-specific configurations are needed to incorporate the needs of multi-biometric systems. Multi-biometric systems employ fusion strategies at different levels in order to increase biometric performance. In this work, we use the feature and decision level fusion of templates from different modalities, 
Figure~\ref{fig:overview} provides a conceptual overview. The proposed AMB-FHE is (i) \emph{sequential} -- Additional biometric samples are requested only if decision threshold is not met, (ii) \emph{cascaded} -- Pass/fail thresholds are applied to determine if additional biometric samples are required to reach a decision, (iii) \emph{decision-level} -- Fusion is performed at the decision level, \ie each comparison outputs its own binary result. In summary, our main contribution is a flexible authentication scheme that stores fused biometric references from different modalities together in a single ciphertext, thereby enhancing privacy protection and efficient utilization of ciphertext slots while preserving run-time flexibility and improved usability.
\par This paper is organized as follows: \ref{sec:related-work} provides a review of related works in the scientific literature. \autoref{sec:proposed-system} presents theoretical background of HE and distance computation and presents the proposed system. In \ref{sec:evaluation}, results of the experimental evaluation are presented. Finally, \ref{sec:conclusion} discusses the conclusions and future work.

\section{Related Work}
\label{sec:related-work}
Homomorphic encryption (HE) is used in various contexts, from genetic analysis and scientific computation to neural network inference. Moreover, HE has been extensively utilized to safeguard various biometric traits. A comprehensive framework for protecting multi-biometric templates using HE was introduced and evaluated with the Paillier cryptosystem in\cite{gomez-barreroMultiBiometricTemplateProtection2017}. Modern fully homomorphic encryption (FHE) schemes leverage \emph{batching} to mitigate computational overhead. Batching encodes multiple plaintexts into a single ciphertext, enabling simultaneous operations on all encoded values (SIMD), thereby enhancing efficiency.

In 2018, Boddeti proposed a secure face matching system\cite{Boddeti-FaceBTP-FHE-BTAS-2018} utilizing FHE based on the Fan-Vercauteren scheme\cite{fanSomewhatPracticalFully2012}, which operates on integers. The system employs batching and quantization to speed up distance computations. The homomorphic inner product is computed through cyclic rotation of the encrypted vector, a technique inspired by Gentry's introduction of slot rotations as a homomorphic operation\cite{gentryFullyHomomorphicEncryption2012}. Experiments demonstrated that while FHE is significantly slower, batching substantially accelerates the comparison process, making the system viable for secure face comparison.

Maheshkumar \etal~\cite{maheshkumarBMIAEBlockchainbasedMultiinstance2020} proposed a blockchain-based multi-instance iris authentication system, applying additive ElGamal encryption to iris templates before blockchain storage to address privacy concerns.

Sperling \etal~\cite{sperlingHEFTHomomorphicallyEncrypted2022a} developed a method for securely fusing and matching biometric templates at the feature level using FHE. It non-interactively performs feature concatenation, fusion, dimensionality reduction, scale normalization, and score computation with an FHE-aware algorithm, overcoming the absence of non-arithmetic operations in HE schemes through learned linear projection. This approach outperformed single biometric representations.

Bauspiess \etal~\cite{bauspiessImprovedHomomorphicallyEncrypted2022} enhanced biometric identification with FHE by storing multiple templates in a single ciphertext, enabling simultaneous comparison of all protected templates, thus reducing the number of expensive homomorphic operations.

In\cite{bauspiessMTPROMultibiometricTemplate2023}, a framework for biometric template protection (BTP) was presented, integrating multi-biometric template protection, FHE, and secure multi-party computation (MPC) for secure and privacy-preserving multi-biometric verification. Homomorphic transciphering was employed to maintain privacy even if both protected templates and the corresponding FHE key are compromised. Singh \etal~\cite{singhSecuringBiometricData2024} evaluated secure matching using FHE on iris and face templates, demonstrating that deep neural network (DNN) extracted face templates fused with dimensionality-reduced iris features yielded optimal performance with the MagFace loss function.

Vallabhadas \etal~\cite{vallabhadasMultimodalBiometricAuthentication2023} presented a multi-modal BTP approach based on the Fan-Vercauteren (BFV) scheme\cite{fanSomewhatPracticalFully2012}, performing feature-level fusion for binary iris and fingerprint templates. The system achieved an equal error rate (EER) of 0.01~\% by compressing, concatenating, and encrypting rotation-invariant templates, followed by Hamming Distance computation.

\section{Proposed System}
\label{sec:proposed-system}
In this section, we present a brief theoretical background on HE, the Cheon-Kim-Kim-Song (CKKS)~\cite{cheonHomomorphicEncryptionArithmetic2016} scheme and the distance computation method used in this work. Finally, we provide a mathematical and algorithmic description of AMB-FHE.
\subsection{Background}
A homomorphism means that operations on the ciphertext yield the same results on the plaintext, thereby allowing for outsourced computation without the need for decryption.
%
\begin{equation}\label{eqn:homomorphic1}
Enc_{pk}(A) + Enc_{pk}(B) = Enc_{pk}(A+B)
\end{equation}
\begin{equation}\label{eqn:homomorphic2}
Enc_{pk}(A) \cdot Enc_{pk}(B) = Enc_{pk}(A \cdot B)
\end{equation}
Usually, HE schemes are categorized by (i) the set of functions they allow to evaluate, (ii) the number of function evaluations, and (iii) the type of input they operate on. Take, for example Equations \ref{eqn:homomorphic1} and \ref{eqn:homomorphic2}, describing the additive and multiplicative homomorphic property, respectively. In this, plaintexts $A$ and $B$ are encrypted using the public key $pk$. Evaluation of a function on the resulting ciphertexts produces a new ciphertext that, upon decryption, reveals the product of $A$ and $B$, allowing for privacy-preserving computation.
Privacy homomorphisms have been known and studied since the invention of public key cryptography in the late 1970s. Multiplicative homomorphism is a well-known property of \eg RSA, which means that it is possible to multiply the ciphertexts without access to the decryption keys, resulting in the correct result after decryption~\cite{acarSurveyHomomorphicEncryption2018}. RSA can therefore be categorized as an early example of \emph{partially} HE schemes (PHE). The Paillier cryptosystem~\cite{paillierPublicKeyCryptosystemsBased1999} extends the functions that can be evaluated to include the addition of two ciphertexts but provides multiplication only with a constant. RSA and Paillier are therefore considered PHE schemes as the number and type of operations are limited.
In 2005, Boneh \etal~\cite{bonehEvaluating2DNFFormulas2005} developed somewhat homomorphic encryption (SWHE), paving the way for the development of the first FHE scheme by Gentry~\cite{gentryFullyHomomorphicEncryption2009} (that is, allowing arbitrary operations). Allowing an arbitrary number of operations was made feasible by introducing the concept of bootstrapping which refreshes the ciphertext. This comes at immense computational cost. Therefore, leveled schemes have been developed to work without bootstrapping. Additionally, schemes operating on data types other than binary have been invented, most notably BGV~\cite{brakerskiFullyHomomorphicEncryption2011} which works on integers. CKKS~\cite{cheonHomomorphicEncryptionArithmetic2016} has gained a lot of interest due to being the first to implement fixed-point arithmetic in the encrypted domain.

Most modern FHE implementations are fundamentally based on the shortest vector problem (SVP) in lattices. Being asymmetric, they solve the problem of key distribution that is inherent to symmetric encryption schemes. 
\subsection{CKKS}
Here, we recall and briefly define the relevant operations of~\cite{cheonHomomorphicEncryptionArithmetic2016}:
\begin{itemize}
    \item $(sk,pk,evk) \leftarrow KeyGen(\lambda)$: given the security threshold $\lambda$, produce a set of secret key $sk$, public key $pk$ and corresponding evaluation keys $evk$, \eg for rotation.
    \item $c_m \leftarrow Enc_{pk}(m)$: using the public key $pk$, encrypt the message $m$ to obtain the ciphertext $c_m$.
    \item $m' \leftarrow Dec_{sk}(c_m)$: on input of the secret key $sk$, decrypt the ciphertext for the message $m'$.
    \item $Eval_{evk}(f, c_{m1},c_{m2})$: given the evaluation key $evk$, compute function $f$ on the ciphertext.
\end{itemize}\footnote{For brevity, we omit encoding and decoding operations.}
Note that Galois keys corresponding to the number of rotations have to be generated and passed to the client explicitly. Noise is (i) introduced during encryption to prevent encrypting plaintext twice to produce the same ciphertext, thereby achieving indistinguishable under chosen plaintext attack (IND-CPA) security, and (ii) during homomorphic evaluations, particularly multiplications. This results in a probabilistic decryption such that $m=m'$ is true with very high probability. Noise management is a characteristic of the scheme chosen and requires that the parameters be chosen application-specific in order to not exceed the noise budget during computation.
\subsection{Distance Computation}
A common distance metric for feature vector comparison is the Squared Euclidean Distance (SED). Since batched FHE schemes do not allow for access of individual feature vector elements and we did not want to miss out on the performance gains, we computed the distance score using the Hadamard product of two feature vectors as proposed by~\cite{Boddeti-FaceBTP-FHE-BTAS-2018}. This metric has proven to work well for features extracted using DNNs and is compatible with schemes implementing SIMD operations.
We implemented sequential cascaded decision level fusion by incrementally computing the dissimilarity score for each stored reference modality and comparing against a predefined threshold associated with a specific false match rate (FMR).
\subsection{Algorithm}
Biometric verification is implemented by incrementally computing the distance between the encrypted concatenated reference and probe templates. This is done using Algorithm~\ref{alg:incremental}. After computing the inner product, the sum of the elements in the ciphertext is calculated by rotation and addition of the ciphertexts. If comparison of the first modality yields a distance $\delta_1$ smaller than the chosen FMR threshold $\tau_1$, the authentication is deemed successful and finished. Otherwise, the user is requested to present the second modality, perform pre-processing, feature extraction, and comparison against the second FMR threshold and so forth. The partial distance calculation was performed by rotating the vector $d$ slots, with $d$ denoting the template size. As such, using public Galois keys, the ciphertext slots can be rotated in order to implement Hadamard product computation.
Following the notation introduced in~\cite{gomez-barreroMultiBiometricTemplateProtection2017}, we use $\mathbf{T}$ to refer to a stored template, with $\mathbf{T}^1$ and $\mathbf{T}^2$ referring to the first and second modalities in a concatenated template. Similarly, we use $\mathbf{P}$ to denote the corresponding \emph{protected} template. Note that the AMB-FHE algorithm can be extened to more than two modalities as described in the following pseudocode.
\begin{algorithm}[H]
 \caption{Pseudocode implementation of AMB-FHE.}
 \label{alg:incremental}
\begin{algorithmic}[1]
\renewcommand{\algorithmicrequire}{\textbf{Input:}}
\renewcommand{\algorithmicensure}{\textbf{Output:}}
 \REQUIRE {Protected probe templates $\mathbf{P}_p^1, \ldots \mathbf{P}_p^m \in \mathbb{R}^d$}, Protected reference template $\mathbf{P}_r^m \in \mathbb{R}^d$, Template length $d$, FMR thresholds $ \{\tau_1, \ldots, \tau_m\}$
 \ENSURE {Decision (accept/reject) $Decision = x \in (0,1)$}
\textit{Initialisation} :
\STATE read database record $\mathbf{P}_r$\;
    \FOR{$j \gets 1$ to $m$}
        \STATE Acquire next probe template $\mathbf{P}_p^j$
        \STATE $\mathbf{P}_r \leftarrow$ $Eval_{evk}(\texttt{Rot}(\mathbf{P}_r, d-1)$)
        \STATE $\delta \gets \delta + $ \texttt{InnerProduct}($\mathbf{P}_p, \mathbf{P}_r^j$)
        \IF{$\delta < \tau_i$}
            \STATE Decision $\gets$ ACCEPT
            \STATE break
        \ENDIF
    \ENDFOR
\STATE Decision $\leftarrow$ REJECT

\end{algorithmic}
\end{algorithm}
\subsection{Protocol}
\begin{figure}[ht]
\centering
\includegraphics[width=.98\linewidth]{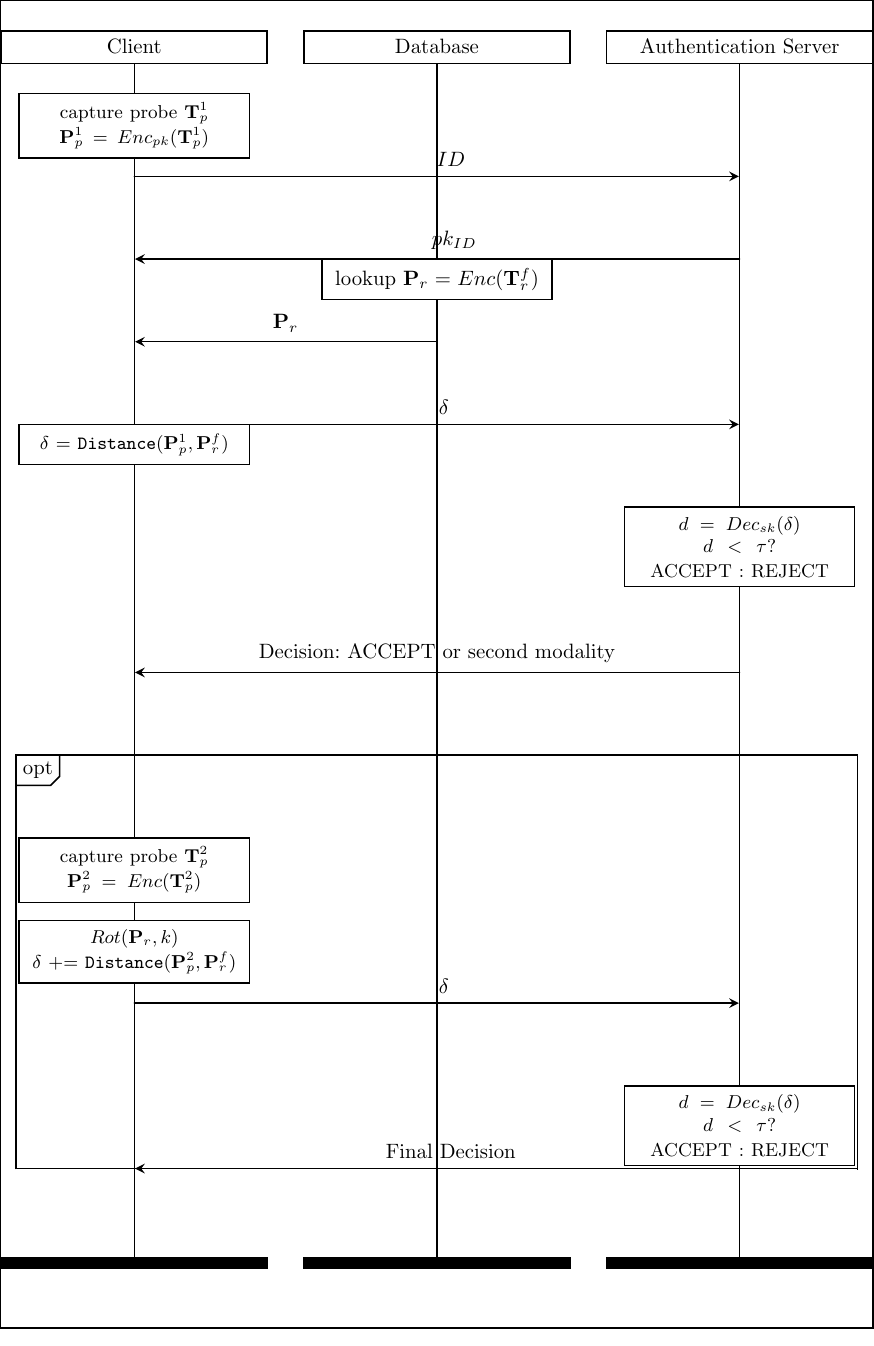}
\caption{Proposed AMB-FHE protocol for authentication based on Model I (Store on server, compare distributed) concept presented in ISO/IEC 24745.}
\label{fig:methodology:protocol}
\end{figure}
In our protocol, aspects of both decision- and feature-level fusion are used.
The protocol shown in Figure~\ref{fig:methodology:protocol} implements sequential cascaded decision-level fusion for two modalities as described in~\cite{ISO-IEC-24722-TR-Fusion-150216}. 
The system implements adaptive multibiometric verification in the encrypted domain. AMB-FHE is executed in the following way:
\begin{enumerate}
        \item Enrolment: The client captures a biometric sample of the first and second modality, performs feature extraction, yields $\mathbf{T}_r^1$ and $\mathbf{T}_r^2$, concatenates them into $\mathbf{T}_r^f$ and encrypts them using the servers public key $Enc_{pk}(\mathbf{T}_r^f)$.
        \item Verification: The client captures a biometric sample of the first modality, performs feature extraction and yields $\mathbf{T}_p^1$. The probe template is encrypted using the servers public key $Enc_{pk}(\mathbf{T}_p^1)$.
        \item The server searches for the encrypted reference template of the claimed identity $\mathbf{T}_r$ and sends it to the client.
        \item The client performs a distance computation in the encrypted domain for the first probe template and sends the ciphertext containing the score $Enc_{pk}(\delta)$ to the server.
        \item The server decrypts the ciphertext, compares the score with the threshold $\tau_1$ and sends the decision (ACCEPT or REJECT) to the client.
        \item If the decision is ACCEPT, the protocol ends successfully. If the decision is REJECT, the client is given another chance to capture a probe of the second modality.
        \item The client captures a biometric sample of the second modality, performs feature extraction, yields $\mathbf{T}_p^2$, encrypts it using the servers public key.
        \item The client performs a slot rotation according to the length of the first reference template and increases the distance score in the homomorphic domain.
        \item The client sends the ciphertext containing the distance score for the first and second modality to the server, $Enc_{pk}(\delta)$.
        \item The server decrypts the ciphertext, compares the score with the threshold $\tau_2$ and sends the final decision (ACCEPT or REJECT) to the client.
\end{enumerate}
\section{Evaluation}
\label{sec:evaluation}
In this section, we present databases, experiments and results.
\subsection{Software and Databases}
The experiments were carried out on a virtual multi-modal database consisting of iris and fingerprint images from the CASIA Iris Thousand ~\cite{casia} and the MCYT fingerprint databases \cite{ortega-garciaMCYTBaselineCorpus2003}, respectively. DNN-based real-valued feature vectors of length 512 were extracted using the method of~\cite{hafner_deep_2021} taken from~\cite{shahrezaBenchmarkingCancelableBiometrics2023} for iris and the method of ~\cite{rohwedderBenchmarkingFixedlengthFingerprint2023} for fingerprints. The database contains fingerprint and iris templates for 533 virtual subjects, with the number of samples per modality varying between 2 and 5. In total, there are 4,584 samples in the database. No explicit normalization has been performed, since the feature vectors have the same dimension and produce distance scores of the same order of magnitude.
\subsection{Performance Evaluation}
\begin{table*}
	\scriptsize
	\centering
	\caption{Relative performance of CKKS operations.}
	\label{tab:relative-performance}
    \begin{adjustbox}{max width=\textwidth}
\begin{tabular}{p{5mm}ccccccccccl}
	\toprule
	$\lambda$ & Parameters & $N$ & \multicolumn{8}{c}{Time (normalized to $Add$)}                                                                                       \\
	\cmidrule{4-12}
	                             &            &             & $Ecd$                        & $Dcd$ & $Enc$ & $Dec$ & $Add$ & $Mul$ & $Rot_1$ & $InnerProduct_{512}$ & $InnerProduct_{1024}$ \\
	\midrule
  128                          & PN12QP109  & 4096        & 10.2                     & 80.4    & 66.8     & 13.6     & 1  & 5.9 & \textbf{90.1}   & 48590 & 99330         \\
  128                          & PN13QP218  & 8192        & 18.8                    & 140.3    & 88.8     & 24.2     & 1  & 5.4 & \textbf{53.1}   & 27884    & 56543        \\
	\bottomrule
\end{tabular}
\end{adjustbox}
\vspace{0.2cm}\newline
    
    $\lambda$, security threshold (bits). $N$, dimension. $Ecd$, Encode. $Dcd$, Decode. $Enc$, Encrypt. $Dec$, Decrypt. $Add$, Addition. $Mul$, Multiplication. $Rot_1$, $InnerProduct_{512}$, Inner Product ($d=512$).$Rot_1$, $InnerProduct_{1024}$, Inner Product ($d=1024$).
\end{table*}
For reference, we implemented baseline unprotected uni-biometric and multi-biometric verification. As metric, the equal error rate (EER) is used, \ie the point at which the false non-match rate (FNMR) is equal to the false match rate (FMR). For the multi-biometric case, fusion occurred at the feature level through concatenation of the vectors. For the adaptive algorithm, we implemented sequential cascaded decision-level fusion with the OR operator. Doing so, we are able to (i) increase security of stored modalities, (ii) make use of the performance benefits of batching, and (iii) avoid capturing modalities that are not needed in case of a good first comparison score below the acceptance threshold.
The values given in Table~\ref{tab:relative-performance} are the relative time requirements for some HE operations of the CKKS scheme. It is evident that operations require vastly different amounts of time, with rotations being especially expensive, accounting for $\approx 96\,\%$ of the total time. To further accelerated CKKS, Bassit~\etal introduced Lookup Table-based Biometric Comparators in order to perform decisions under encryption~\cite{bassitMultiplicationFreeBiometricRecognition2022}.

One of the reason for the popularity of HE schemes is the fact that their impact on accuracy is very small. In our evaluation, average and maximum score deviation for mated and non-mated comparisons are smaller than $10^{-4}$ even in the worst case, biometric performance is unimpaired compared to the unprotected system.
%
Single biometric fingerprint and iris systems perform similar, with EER at $1.23\,\%$ and $1.20\,\%$, thus iris performing slightly better. Fusion of both modalities results in an EER of $0.08\,\%$.
Whilst parallel decision-level fusion in the encrypted domain is easy to implement, we have shown that sequential capturing of biometric data and decision-level fusion in the homomorphic domain is feasible as well and can be sped up using slot rotations. This reduces the computational overhead from $\mathcal{O} \approx (\frac{d}{2}+d)$ to $\mathcal{O} \approx d+k$, where $d$ is the feature vector length and $k \ll d$ a negligible parameter-dependent constant overhead of rotation and decryption. This is a nice speed up compared to the na"ive approach of re-computing the comparison of both fused feature vectors.
The following systems have been implemented:
\begin{itemize}
    \item AMB-FHE-1 (OR, iris first)
    \item AMB-FHE-2 (OR, fingerprint first)
    \item Multi (AND)
\end{itemize}

\subsection{Usability}
The computational performance gains directly and indirectly translate to usability improvements due to (i) faster verification times and reduced energy consumption and (ii) saved modality presentations. We used the unconditionally multi-biometric system as the baseline for comparison. These improvements are presented in \autoref{tab:results:saved_presentations} at FMR thresholds 0.01\,\%, 0.1\,\% and 1\,\% for the adaptive case. Depending on the FMR threshold set, saved 2nd modality presentations range from 72\% to 96\% for FMR thresholds between 0.01\% and 1\%. A higher FMR corresponds to a higher distance threshold and therefore lower system security, \ie the system is more likely to accept an impostor. Data for this metric was generated by computing only mated comparison scores as this is (i) the predominant use case and (ii) we do not care for usability in the impostor case.
\begin{table}[ht]
        \centering
        \caption{Number of modality presentations for unconditional fusion and AMB-FHE.}
        \label{tab:results:saved_presentations}
        \begin{tabular}{ccccc}
                \toprule
                System    & Threshold (\%) & Iris & Fingerprint & Saved (\%)     \\
                \midrule
                Multi    & -              & 1296 & 1296        & -              \\
                AMB-FHE-1 & FMR=0.01       & 1296 & 360         & 936 (72.22\%)  \\
                AMB-FHE-1 & FMR=0.1        & 1296 & 247         & 1049 (81.02\%) \\
                AMB-FHE-1 & FMR=1          & 1296 & 51          & 1347 (96.1\%)  \\
                AMB-FHE-2 & FMR=0.01       & 338  & 1296        & 958 (73.9 \%)  \\
                AMB-FHE-2 & FMR=0.1        & 221  & 1296        & 1075 (82.9 \%) \\
                AMB-FHE-2 & FMR=1          & 48   & 1296        & 1344 (96.3 \%) \\
                \bottomrule
        \end{tabular}
\end{table}
\subsection{Security Analysis}
In this section, we provide security analysis according to the requirements imposed by ISO/IEC 24745,  (i) \emph{irreversibility}: it is impossible to construct valid samples from protected templates, (ii) \emph{unlinkability}: an attacker must not be able to link two encrypted templates from different applications to the same subject, (iii) \emph{renewability}: it shall be possible to create new templates from an enrolled subject that is different from the old templates.
Due to their probabilistic nature, encrypting the same plaintext twice results in different ciphertexts, \ie templates as well as computed distance scores fulfill the \emph{unlinkability} and \emph{renewability} requirement.
Given that the security guarantees of the underlying cryptographic primitives hold, \emph{irreversibility} is also fulfilled under the assumption that an attacker is not in possession of the private key.
Storing multiple biometric templates together improves the security against \emph{offline} attacks. However, if the decision-level fusion function was chosen to be OR (as we did), online attacks on arbitrary biometric modalities inside the ciphertext are possible as a side-effect of better usability. That is, an attacker is able to purposefully fail the first modality and attempt the following ones in order to (i) gain access or (ii) learn information about the stored templates by observing the binary decision. Being an online attack, implementing mitigations at the protocol level such as rate and retry limiting is possible and effective. It is, however, very important that the authentication server only returns a binary decision in order to prevent effective Hill-Climbing attacks. For more information on attack paths against parallel and sequential multi-biometric systems, we refer the reader to the work of Maiorana~\etal~\cite{maiorana_hill-climbing_2015}.
If the attack paths discussed above are viewed as an online attack on the authentication system, the privacy aspect can be interpreted as an offline attack on the stored templates. While measures such as retry and rate-limiting are easy to implement and effective for online attacks in order to fend them off despite a low theoretical security level (FMR $=0.1\% corresponds to, 50\%$ probability with only $\sqrt{1000} \approx 32$ attempts), one must also consider recovering protected templates thereby violating privacy as an attack vector. As even protected multi-biometric templates are insufficient to reach relevant security levels of $\approx 128$ bits~\cite{rathgebMultiBiometricFuzzyVault2023}, protecting against this attack vector requires additional secrets such as PIN or passwords, which are, however, less convenient to use.
\section{Conclusion and Outlook}
\label{sec:conclusion}
In this work, we presented an approach to securely store multiple concatenated biometric references and perform partial comparison in the encrypted domain, thereby achieving authentication with flexible security levels chosen at run-time and usability improvements. We have shown the potential of feature-level fusion \wrt HE.
The idea of incremental matching~\cite{yiFastMatching2013} is particularly useful in the encrypted domain and can be applied \emph{within} a template of a given modality, achieving even more control over the security and performance characteristics. The impact can be greatly increased by having feature vector values sorted by priority if this is applicable for a given modality.
HE has proven to be a valuable tool for addressing biometric template protection, and thus we point out the need to further improve efficiency.
The work presented is limited to fixed-length representations of the same data type. Preliminary work on comparison of variable-length templates has been done in~\cite{gomez-barreroPrivacyPreservingComparisonVariableLength2017}.
This needs adaption of responsibilities or verifiable computation and remains an open problem.
Applying quantization after feature extraction and the use of HE schemes working on integers or binary data showed large gains in computational complexity with moderate impact on biometric performance~\cite{drozdowskiBenchmarkingBinarisationSchemes2018}. However, the CKKS scheme chosen in this allows one to mix data types at the cost of high computational workload.
\section*{Acknowledgment}
This research work has been funded by the German Federal Ministry of Education and Research and the Hessian Ministry of Higher Education, Research, Science and the Arts within their joint support of the National Research Center for Applied Cybersecurity ATHENE.
\bibliographystyle{IEEEtran}
\bibliography{references}
\end{document}